\documentclass{article}

\usepackage{arxiv}

\usepackage[utf8]{inputenc} % allow utf-8 input
\usepackage[T1]{fontenc}    % use 8-bit T1 fonts
\usepackage{hyperref}       % hyperlinks
\usepackage{url}            % simple URL typesetting
\usepackage{booktabs}       % professional-quality tables
\usepackage{amsfonts}       % blackboard math symbols
\usepackage{nicefrac}       % compact symbols for 1/2, etc.
\usepackage{microtype}      % microtypography
\usepackage{lipsum}		% Can be removed after putting your text content
\usepackage{graphicx}
\usepackage{natbib}
\usepackage{doi}

\title{Quoka Atlas of Scholarly Knowledge Production: An Interactive Sensemaking Tool for Exploring the Outputs of Research Institutions}

\author{ \href{https://orcid.org/0000-0002-1657-9809}{\includegraphics[scale=0.06]{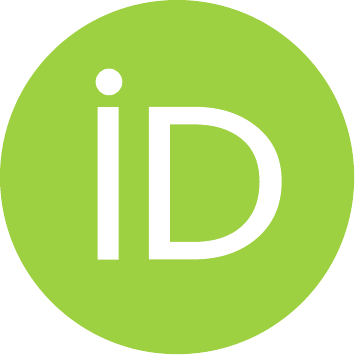}\hspace{1mm}Benjamin Adams} \\
Department of Computer Science and Software Engineering \\ 
University of Canterbury \\ 
Christchurch 8140, New Zealand \\
\texttt{benjamin.adams@canterbury.ac.nz} \\
	\And
	\href{https://orcid.org/0000-0001-8288-5241}{\includegraphics[scale=0.06]{orcid.pdf}\hspace{1mm}Richard Hosking} \\
	Centre for Culture and Technology \\ 
	Curtin University \\
	Bentley 6102, Western Australia \\
	\texttt{richard.hosking@curtin.edu.au} \\
}

\date{}

\renewcommand{\shorttitle}{Interactive atlas of scholarly knowledge production}

\begin{document}
\maketitle

\begin{abstract}
	The vast amount of research produced at institutions world-wide is extremely diverse, and coarse-grained quantitative measures of impact often obscure the individual contributions of these institutions to specific research fields and topics. We show that by applying an information retrieval model to index research articles which are faceted by institution and time, we can develop tools to rank institutions given a keyword query. We present an interactive atlas, Quoka, designed to enable a user to explore these rankings contextually by geography and over time. Through a set of use cases we demonstrate that the atlas can be used to perform sensemaking tasks to learn and collect information about the relationships between institutions and scholarly knowledge production.
\end{abstract}

% keywords can be removed
\keywords{research evaluation \and university ranking \and scientometrics \and sociology of knowledge \and atlas \and information retrieval}

\section{Introduction}

Metrics derived from publication outputs are often a key factor in published quantitative rankings of universities and other research institutions, but these rankings are derived using coarse-grained measures that do not enable an observer to gain insight into the unique scholarly contributions of an institution \citep{huang2020comparison}. Thus, ultimately they are not fit for the purpose of understanding the capacity or track record of a university or department to support a specific line of research or project. Likewise, they do not provide historical context of the institution's role in developing fields of research. Yet without an alternative, these rankings are often used as indicators of a ``good choice'' for prospective students, funding agencies, and national review boards. The principle of particularity proposed by \citet{marginson2014university} is that rankings should not rely on proxy measures that do not directly measure the particular qualities of universities that they purport to capture. Certainly, aggregate rankings fail that standard for anything but the most basic questions of relative merit or reputation \citep{collins2016ranking}. If we wish to understand the distinctive qualities of each university's research contributions then we need better ways to ask finely-honed questions about what those universities are producing now and have produced over time. 

In fact, information that could enable us to answer those more nuanced questions does exist within the abstracts and texts of the publications that the researchers at an institution generate. Thus, it would advantageous to create techniques that allow us to directly tap into that information. Indeed, information retrieval methods that have been developed for ranking web search results provide us with a flexible set of approaches to explore the bespoke qualities of universities' research contributions \citep{white2007combining,bar2016bibliometrics}. Providing the ability for end users' to define ad hoc keyword queries and retrieve ranked results from an index opens the door to millions of different ways of exploring the scholarly production of an institution. For example, if we imagine a prospective graduate student wanting to study ``quantum computing'', she might try to find a university that is highly ranked in Computer Science or Physics. Yet this is no guarantee that there are professors or research groups there working on specific problems related to quantum computing that she is interested in. Furthermore, if she decided to change her search to a more refined topic, such as ``quantum decoherence'' the ranked categories are even less useful. Ideally, if she were to look at the number and quality of relevant research publications on quantum computing that are produced by different universities, then that would provide a more appropriate initial indicator of the suitability the institution for her study. Given the sheer number of articles available, and with no way of exploring them by institution, it would be hard to know where to start. Solving this then becomes an information search and, ultimately, a sensemaking problem.

In this article we present the design and implementation of \textsc{Quoka} (or QUesting Open Knowledge Atlas), a data-driven, interactive, and searchable atlas of scholarly knowledge production built from the text and metadata %(titles, abstracts, and semantic labels)
of 100 million research artifacts. The atlas is designed to enable a user to enter an ad hoc keyword query and then explore relevant research outputs of thousands of institutions world-wide over time. In doing so it provides multiple types of feedback on an interactive map and timeline display. At the global level a heatmap display shows geographic regions where relevant publications are being produced. At a more zoomed-in level, individual institutions are visually ranked by markers that are sized based on an information retrieval ranking of the institution's research outputs and depending on the user's query. The timeline display shows changes in the prominence of the topic over time and it also serves as a mechanism to filter the map display by year. Thus, for any given topic, geographic regions and institutions can be compared over different time periods. Additionally, by selecting an institution on the map, the top most relevant publications are shown for the filtered time period. 

The four main contributions of the research detailed in this article are as follows:

\begin{enumerate}
    \item We introduce a technique that organizes and indexes scholarly publications along two dimensions: institution and year of publication, which together provide a useful frame of reference for exploring, searching, and comparing research topics and institutions. Because a text index underlies the proposed system, the user can easily drill down and read specific documents produced at an institution.
    \item While some prior work has been done on geographically and temporally organizing research publications from specific publishers (cf. \citet{gao2013spatiotemporal}), the scale of the data used in \textsc{Quoka}---with regard to both the breadth of topics as well as the number of objects ({\textgreater}100 million)---far exceeds anything done previously; the index is comparable in size to commercial scholarly search engines such as Google Scholar. The result is an open-ended and regularly updated atlas that can be used in multiple disciplinary contexts.
    \item We re-frame the ranking problem as one of sensemaking, whereby the goal is essentially one of schematizing the evidence from a diverse set of data (i.e., research publications) \citep{pirolli2005sensemaking}. For that to work effectively we need a tool that allows a user to forage and collect contextually-relevant information, which is functionality that we prototype in the interactive atlas.
    \item We demonstrate through use cases how the interactive atlas can provide a nuanced historical and institution-based perspective of scholarly production for a set of example topics, and discuss the implications for the sociological study of knowledge production.
\end{enumerate}

The remainder of this article is organized as follows. After providing background material on scientometrics, sensemaking, and information retrieval, we explain the design goals and issues for the atlas, including data collection and preparation. We follow with a description of the implementation of the \textsc{Quoka} system, including both the back-end index as well as the front-end user interface. Finally, we conclude with a brief discussion of the next steps. %in Section 5 we discuss some exemplary use cases for the atlas, and then we end with a discussion of implications for scientometrics and institutional rankings.

\section{Related Work}

Despite the common goal of providing a ranking result based on content, the most prominent ranking methods in science do not incorporate methods from information retrieval \citep{bar2016bibliometrics,mayr2015scientometrics,white2007combining}. Where the academic outputs do play a role in rankings (of either individuals or institutions), it is more often based on factors such as the reputation of the researchers \citep{safon2013global}, the quantity of citations \citep{hirsch2007does}, or other network based measures \citep{ding2009pagerank}. Indeed, rankings based on these measures have been widely critiqued for not capturing the quality of research that is produced and even being harmful to institutions \citep{bornmann2007we,amsler2012university,lynch2015control,biagioli2020gaming}. The goal for what we propose here is not to create a new ranking for the purpose of globally comparing universities, but rather to create tools that can highlight the variety and comparative focuses of many different institutions. For this, the ability to search and contextualize the research outputs of an institution is key.

\subsection{Sensemaking}

Sensemaking is the process that an individual or group engages in to synthesize knowledge in order to aid decision-making from complex and varied evidence \citep{russell1993cost}. In human-computer interaction studies, sensemaking has been framed in terms of constructing representational schemas that explain the evidence, for example, to provide a summary report that synthesizes from data. Sensemaking has been studied in many different domains, including intelligence analysis, medical decision making, and education \citep{pirolliRussell2011}. \cite{pirolli2005sensemaking} have created a model of sensemaking that consists of two major loops, a foraging loop and a sensemaking loop. In the foraging loop information is found and organized into evidence. In the sensemaking loop a structured story is built from this evidence by representing it in a schematic form (e.g., a visualization or narrative) and by creating hypotheses that support the sensemaker's conclusions. The two loops feed into each other, so the creation of a schematic, for example, can lead to a re-evaluation of the evidence or require seeking out additional information. Sensemaking, therefore, is an iterative process that flows in both a bottom-up and top-down manner \citep{pirolli2005sensemaking}. 

Sensemaking has taken on a slightly different definition when viewed from a psychological perspective. It has been characterized as an active form of situational awareness, where frames, or mental models, are created from data and iteratively re-framed to interpret past events and predict future ones \citep{klein2006making,klein2006making2}. Regardless of the formalization used, sensemaking tools are designed to facilitate the iterative steps required in order for someone to come to a conclusion or decision in the face of heterogeneous, often ambiguous, and sometimes contradictory data \citep{kirschner2012visualizing}. Computational tools aid humans who are performing sensemaking by supporting information seeking and foraging tasks, or aiding it through collaborative sensemaking interfaces.

\subsection{Why discovering relationships between institutions and knowledge production is sensemaking}

In this section we describe three example scenarios that demonstrate how coming to an understanding of the scholarly output of an institution or research organization is, in fact, a kind of sensemaking activity as described above. In each case the sensemaker utilizes a combination of bottom-up and top-down processes to reach a conclusion on a problem they are interested in. As described in \cite{pirolli2005sensemaking} the bottom-up processes include searching and filtering for information, reading and extracting evidence from the information that is found, representing the information in a schematic form, such as a narrative, and making a decision. The top-down processes include re-evaluating the conclusions based on external feedback, which will lead to searching for additional support, evidence, and relations to either bolster the previous conclusions or re-evaluate them.

\subsubsection{Scenario 1: A student researching where to go to graduate school.}

In the introduction we presented a scenario of a student researching where to go to for graduate school. Attracting students is one of the common motivators for university rankings, but for students wishing to explore deeper into the offerings at different universities rankings are not particularly useful. Undoubtedly, many potential Ph.D. students have a general interest in a domain of research but they might not have been exposed to the breadth of topics and projects in progress at various universities. It is also possible that there is relevant research being conducted in a different type of department than the one the student is considering, as often occurs in interdisciplinary contexts. Finding a good potential program and advisor is a sensemaking task because it involves distilling the many different forms of evidence through an iterative process whereby the student learns more about the research that is being conducted and who is doing it. This might begin in a bottom-up way with a broad exploratory search through different programs that are available, or in a more top-down manner, where, for example, the undergraduate research advisor has pre-disposed the student to look at a few specific programs, which then expands to other programs as she reads more of the research that is happening there and elsewhere. As the process progresses the student might refine her schema of what type of program is suitable and might even change the focus of her planned research. This can, of course, also change after she has begun her studies, but the decision about where to apply and enrol represents a clear example of sensemaking from heterogeneous information with the goal of making a critical decision that can have significant implications for the students' future. 

%It is common that a student who is planning to study a Ph.D. will have been introduced to a field of study or topic while studying as an undergraduate, perhaps in a course or by conducting a capstone independent research project. However, thought the student might be familiar with the general topic, she might not know what universities specialize in that area or what the current projects are that researchers at those universities are working on. Furthermore, there might be researchers who are working on topics that would be of interest to the student, though they do exist on their radar of universities. There are several information sources that can help in such a case. If it exists, then network connections and expert knowledge at the undergraduate university can be useful. 

\subsubsection{Scenario 2: A national review board wanting to get a picture of the research landscape.} 

Universities and individual researchers are increasingly required to report on the performance of their research output to national review boards \citep{hicks2012performance}. When measuring research performance several factors can be used to measure impact, complicated by the many ways that different disciplines and individual researchers contribute to knowledge production. 

\subsubsection{Scenario 3: A researcher studying the history and sociology of an academic research field.}

Over time new academic research disciplines are created and others shift in focus and scope over time. These developments, as well as understanding the underlying causes and actors involved, are of interest to historians of the academy. Younger research fields, such as geographic information science (GIScience), often have fervent debates about the nature of the discipline \citep{reitsma2013revisiting}, while older disciplines might re-brand or splinter as the field progresses. One example of this is the clear shift in 2004 away from ``humanities computing'' to ``digital humanities'' \citep{hockey2004history,vanhoutte2013gates}. The role of geography in the development of research and innovation has also been extensively studied \citep{saxenian1996regional,jons2013global,Frenken2020}.

\subsection{Chronotopic information interaction}

Chronotopic information interaction is a design paradigm that uses the inherent spatio-temporal structure found in a  heterogeneous document collection to support information seeking behavior. This structure can be derived from document metadata as well as the references to places and dates within the text of the documents \citep{Adams2020}. A search engine that uses this form of interaction visually emplaces search results within an integrated geographical and temporal frame of reference, which provides context to explore and discover information. This geographical and historical context allows a user to leverage their individual expert knowledge about different locations and historical events while exploring the search results \citep{duggan2008knowledge}. When applied to research publications, chronotopic information interaction enables the assessment of an institution's historical role in the development of research as well as a way to explore the contributions that different institutions make over time. In the \textsc{Quoka} system chronotopic information interaction is used to augment sensemaking tasks to understand the role of institutions and researchers in knowledge production over space and time. 

%``Mapping the backbone of science'' \citep{boyack2005mapping}.

\section{Overview of the Atlas}
\begin{figure}
\includegraphics[width=\hsize]{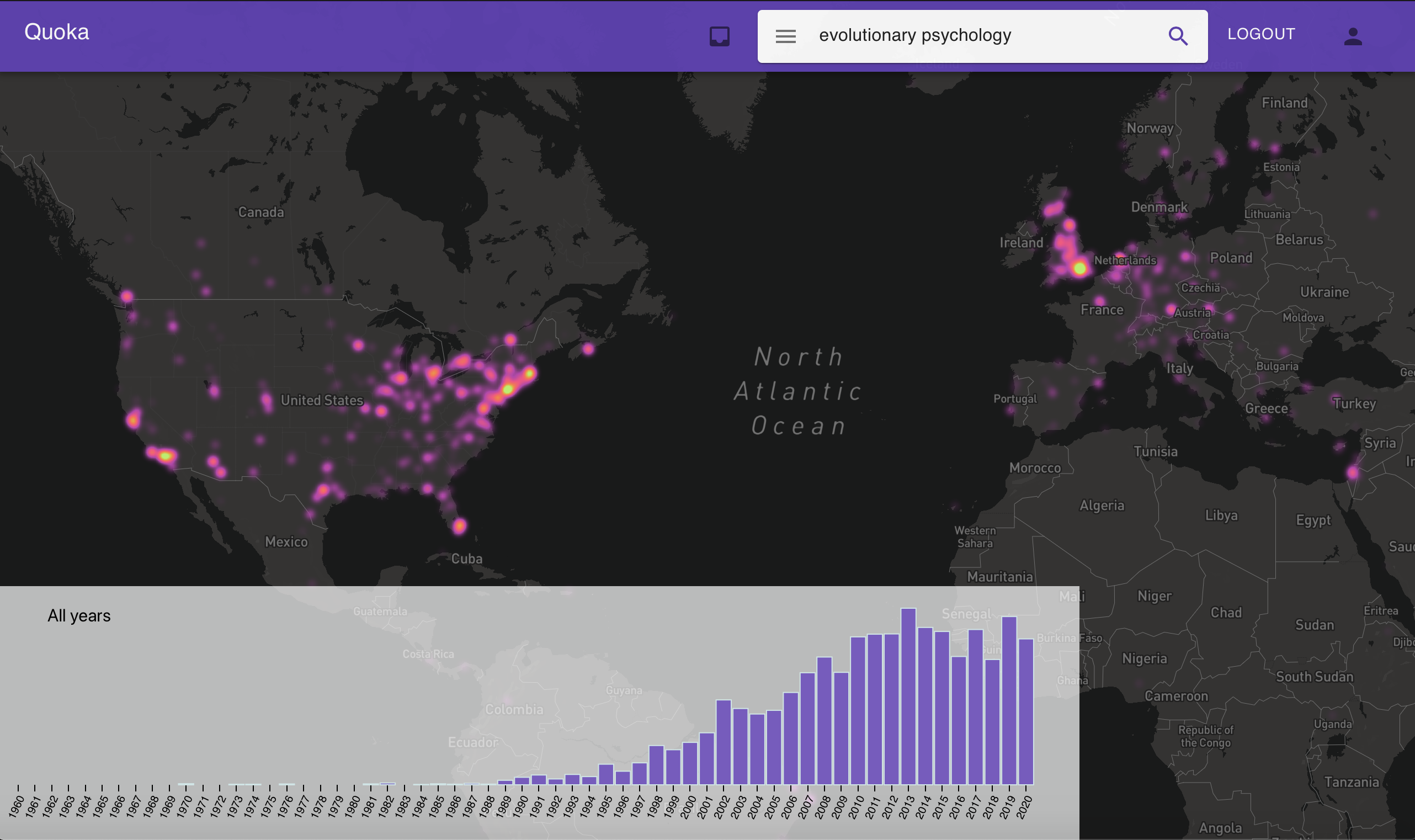}
\caption{Overview of `evolutionary psychology' research in Europe and the United States. The timeline shows that research in this field started slowly in the late 1980s growing in the following decades and then plateauing in the 2010s.}
\label{fig:search_overview}
\end{figure}

The \textsc{Quoka} atlas is a publicly-available interactive website \footnote{\url{https://pteraform.csse.canterbury.ac.nz/quoka/}}. Similar to a standard search engine, a keyword query is entered in a search box and a geographic and historical overview is presented on a web map and timeline as seen in Figure~\ref{fig:search_overview}. In that example the results for the query ``evolutionary psychology'' are shown. A heatmap is rendered on the map derived from the index scores for each institution, and the timeline shows the relative salience of the research topic over time from 1960 to 2020. Since geographically proximate institutions blend together when zoomed out, the heatmap enables one to see geographic clustering of research activity.

\begin{figure}
\includegraphics[width=\hsize]{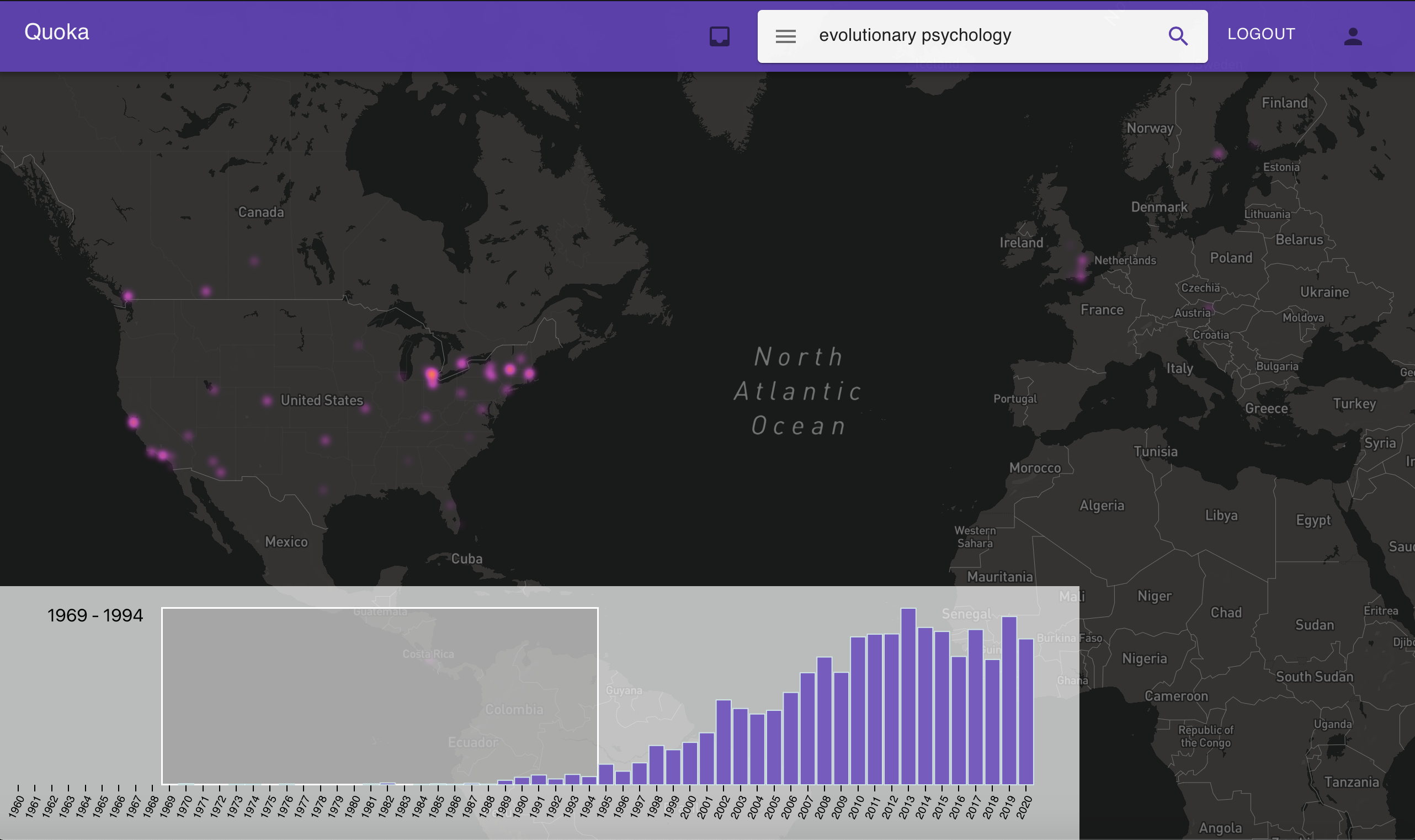}
\caption{Overview of `evolutionary psychology' research 1969--1984.}
\label{fig:search_time_select}
\end{figure}

\begin{figure}
\includegraphics[width=\hsize]{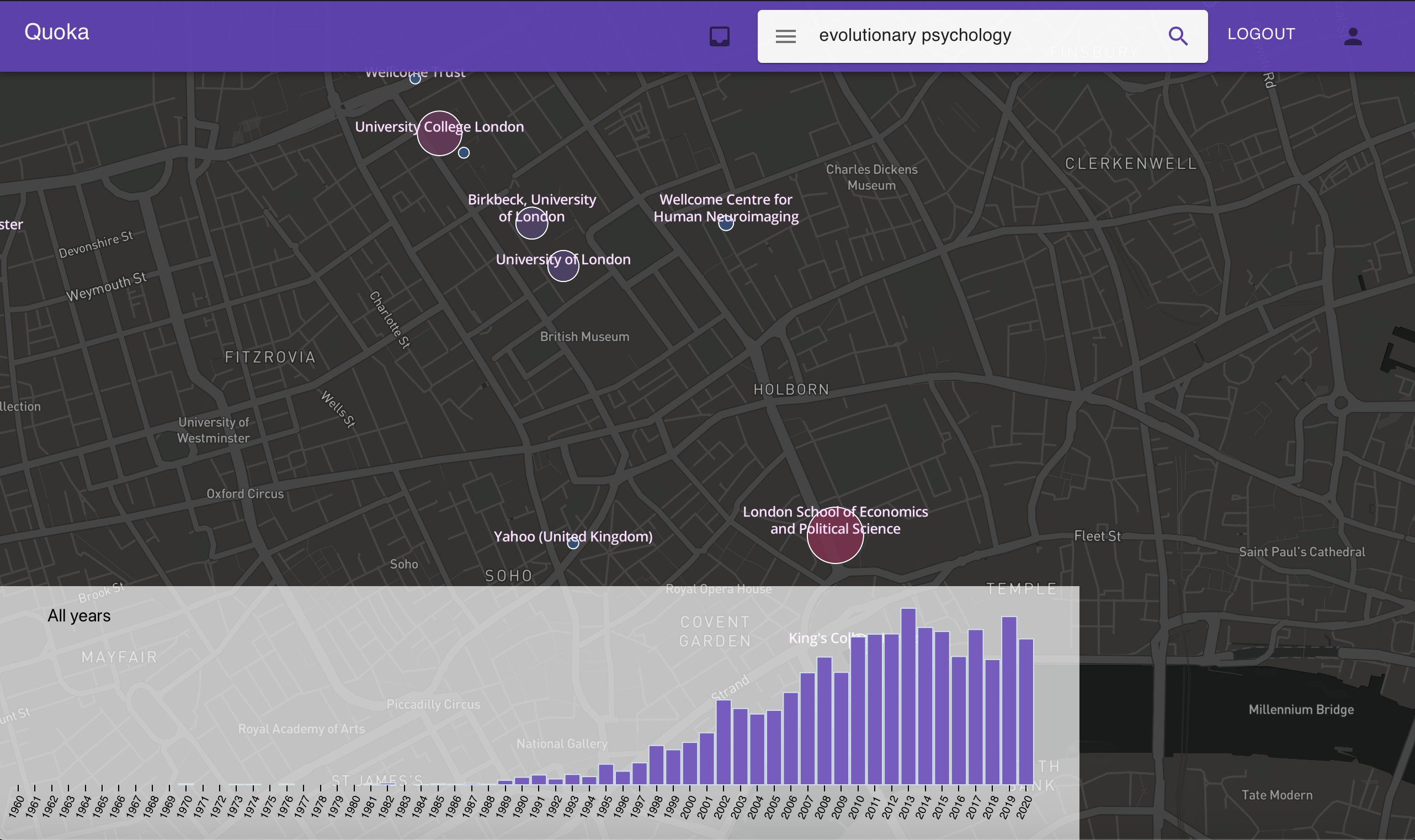}
\caption{Individual institutions and their scores are rendered visually using weighted circles as the user zooms in on the map. Here we see a number of institutions in  London that have produced research related to `evolutionary psychology'.}
\label{fig:search_zoomed}
\end{figure}

\begin{figure}
\includegraphics[width=\hsize]{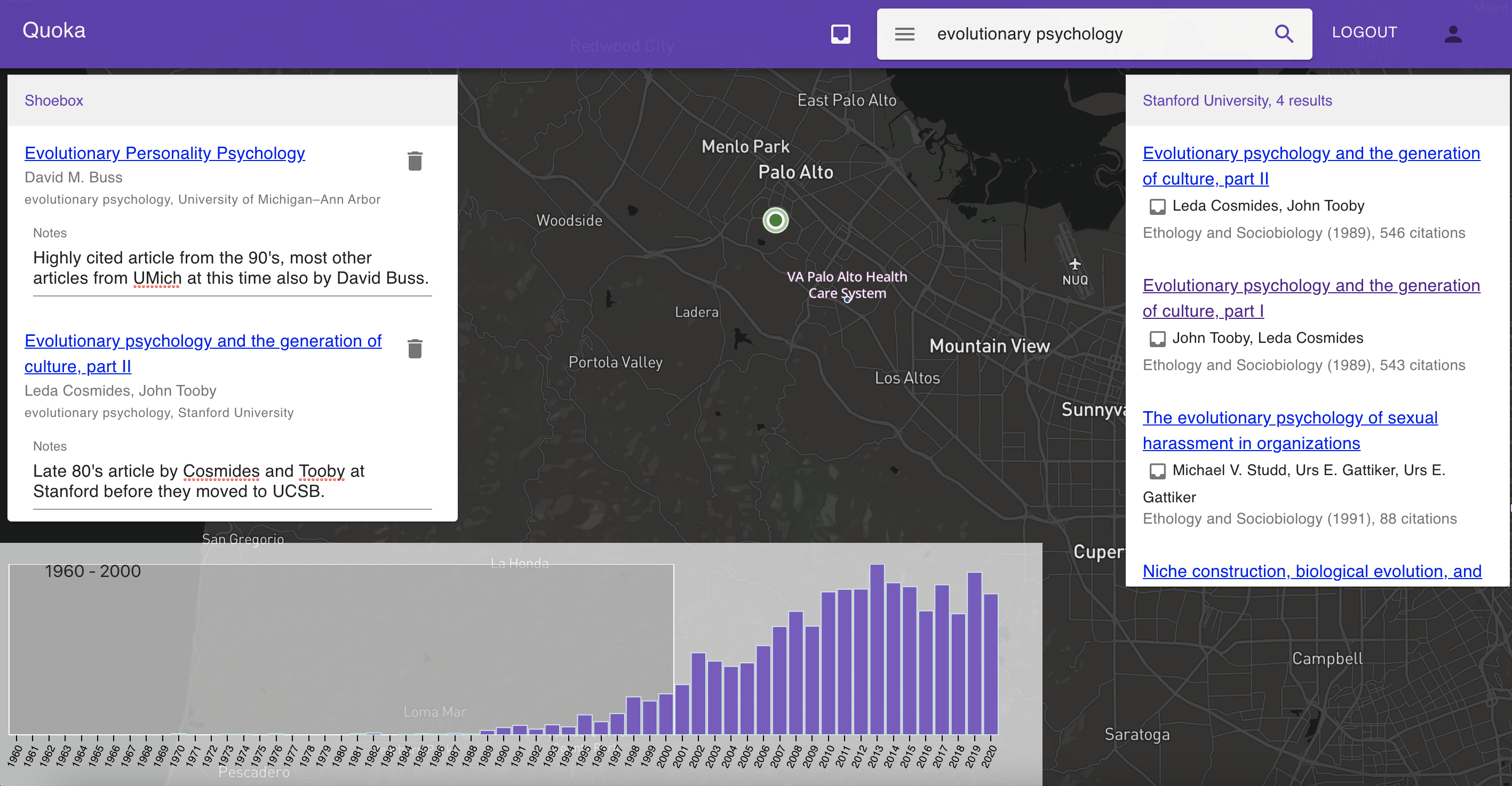}
\caption{Zoom in on Stanford University for `evolutionary psychology' research until 2000. The Shoebox shows two saved articles with notes.}
\label{fig:search_stanford}
\end{figure}

A range of years can be selected along the timeline and the heatmap will adjust accordingly. This allows the user to investigate how the geography of research has changed over time. In Figure~\ref{fig:search_time_select} we see how the field of evolutionary psychology was much less geographically distributed before 1984. As we zoom in on the map, the heatmap fades and individual institutions are marked with sized circles based on the institution's score (see Figure~\ref{fig:search_zoomed}). By selecting an institution with a mouse click we can explore relevant research articles that were published by researchers there. Figure~\ref{fig:search_stanford} shows as we zoom into Stanford University and select it that a ranked list of articles related to evolutionary psychology is presented, filtered by the time range currently selected. 

In their investigation of sensemaking by intelligence analysts, \cite{pirolli2005sensemaking} described how during the initial information foraging steps that lead to sensemaking they made use of a ``shoebox'', an intermediate place to store found information that would be used later to identify potentially relevant relations and evidence, which are subsequently then used to support findings and conclusions \citep{pirolli2005sensemaking}. Being able to share this found information is useful for collaborative sensemaking as well, but additional information which enables others to understand the task context during a search process is important to communicate when handing off the information to others \citep{paul2009cosense}. Therefore, in the shoebox we save the search state (query and institution) that led to an article being found, plus provide a means for the searcher to record additional notes and comments. Figure~\ref{fig:search_stanford} shows that the user has previously added an article from the University of Michigan and has written some notes. 

\section{Implementation}
\subsection{Data sources}
\begin{figure}
\includegraphics[width=\hsize]{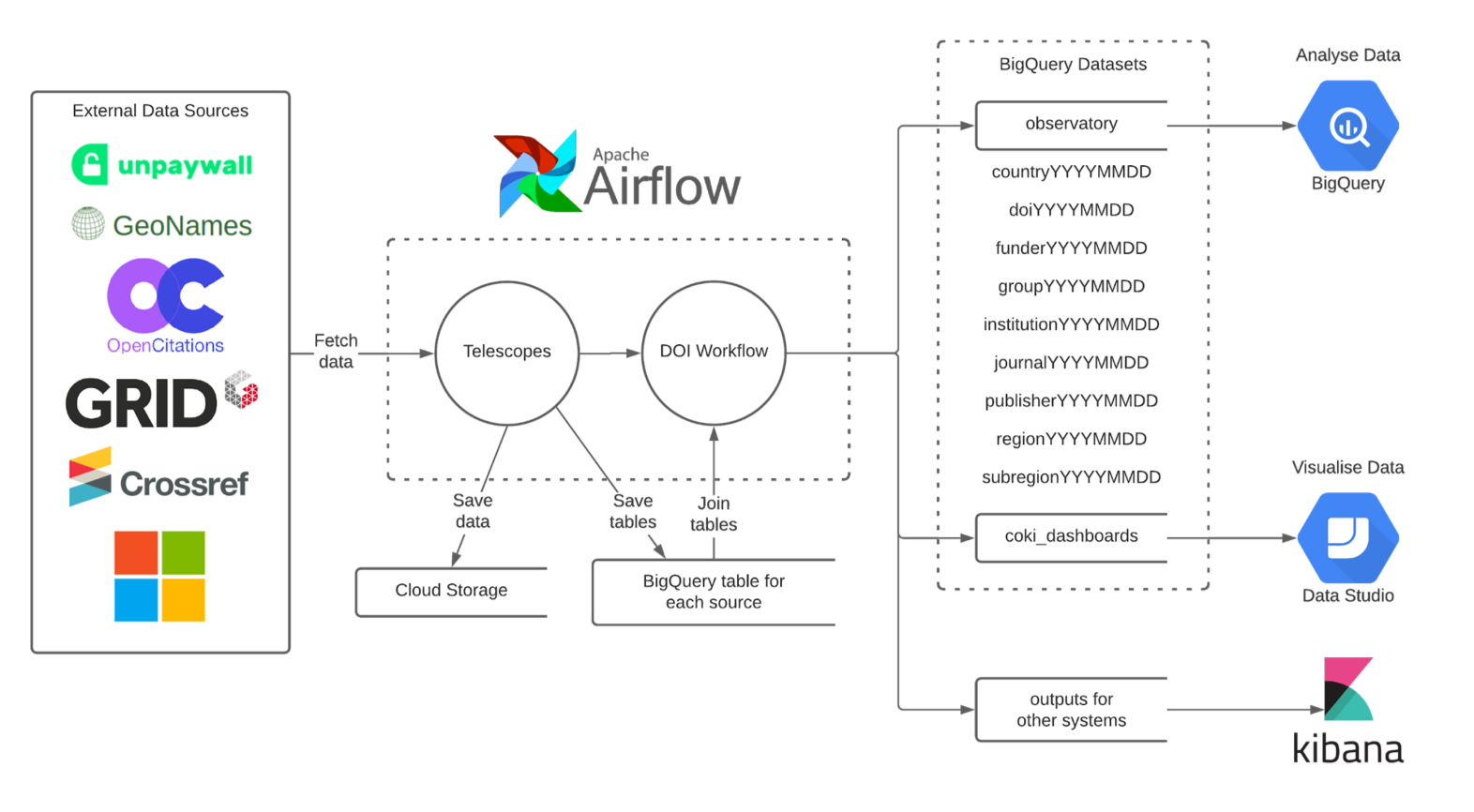}
\caption{COKI Academic Observatory data collection, analysis, and visualization pipeline.}
\label{fig:coki}
\end{figure}

The \textsc{Quoka} indexes are built using the COKI Academic Observatory data collection pipeline, which fetches data about research publications from multiple sources and exposes synthesized data as a collection of Google BigQuery datasets. Figure~\ref{fig:coki} shows a high level schematic of the automated data pipeline and its constituent technologies. Data on over 100 million research artifacts has been collected from Unpaywall, Microsoft Academic Graph, Open Citations, and CrossRef, and this data is updated in an automated manner on a regular basis. Metadata about institutions is matched to this data from Geonames and the Global Research Identifier Database. Exported snapshots of this integrated data power various dashboards and visualizations, and they also serve as the input data for the \textsc{Quoka} indexes.

\subsection{Index creation}
The server back end of \textsc{Quoka} relies on two text indexes built from the Academic Observatory data and are implemented using Elasticsearch and the Apache Lucene library software \citep{mccandless2010lucene,gormley2015elasticsearch}. The current implementation is based on a snapshot of the COKI data from November 2020, and the indexes are in total approximately 400 GB in size.

The first is an index of the aggregate text produced by each institution per year (the \texttt{institution-year index}). This index generates a score for each institution, given a keyword query. These scores can be used for ranking as well as visual feedback, and in the case of the atlas as indicators on the interactive map. The second, the \texttt{DOI index}, is an index of each research artifact organized by digital object identifier. The fields that are stored in the index for each DOI include: title, abstract, Microsoft academic graph fields of study \citep{WangEtAl2020}, authors, publisher, journal name, Global Research Identifier Database\footnote{\url{https://grid.ac/}} (GRID) ids, year of publication, citation count, and open access information. This index allows for ranking of research objects, filtered by institution, year, and other criteria.

\textsc{Quoka} uses an  information-based scoring scheme for both indexes \citep{clinchant2010information}. Given query $q$, the score $S$ for institution, $\alpha$, for a set of years $Y = {y_1, y_2, \dots, y_n}$ is the sum of scores over each $y$ (Equation \ref{eq:map-score}). 

\begin{equation}
  S(q,\alpha,Y) = \sum_{y \in Y} s(q,\alpha,y)
  \label{eq:map-score}
\end{equation}

The institution's score for each year, $y$, is defined as follows in Equation~\ref{eq:iy-score}. 

\begin{equation}
  s(q,\alpha,y) = \sum_{w \in q \cap \alpha,y} -\log \left( \frac{\lambda_{w}^{\frac{t_{\alpha,y}}{t_{\alpha,y} + 1}} - \lambda_{w}}{1 - \lambda_{w}} \right)
  \label{eq:iy-score}
\end{equation}

$t_{\alpha,y}$ is an H2 term frequency normalized version of the sum of occurrences of the word $w$ in the all the text produced by institution $\alpha$ in year $y$ \cite{amati2002probabilistic}. This normalization is based on the number of total words produced by the institution within the year, so that institutions which are producing less content overall though proportionately high with regard to the given query are not scored artificially low. $\lambda_{w}$ is the average number of institution, year pairs where the word $w$ occurs. 

For the index of DOIs, a similar measure is used with an additional normalization parameter for the overall length of the text we have associated with the DOI (title, abstract, etc.). In addition to the text, the year and institutions that contributed to the content are stored in separate fields. This is to enable the retrieval of DOIs for a given institution and filtering to get relevant DOIs that match a given year range.

\subsection{Web application}

The atlas is a single page web application designed to run in modern browser systems. The user interface is built using the React\footnote{\url{https://reactjs.org/}} JavaScript library. The components of the application are constructed using a number of open source software libraries, including Mapbox GL JS\footnote{\url{https://github.com/mapbox/mapbox-gl-js}} for the world map component and D3.js\footnote{\url{https://github.com/d3/d3}} for the timeline display. 

A reactive web server written using Eclipse Vert.X\footnote{\url{https://vertx.io/}} acts as middleware between the Elasticsearch index and the client web application by filtering and handling query requests from the client, and formatting the results as a JSON object for the client application. The use of Elasticsearch and Vert.X technologies provides a scalable architecture that can respond to changing demands on the system from the users.

%\section{Discussion}
%[Go into some in depth examples of the use of the atlas.]
%Uses: comparison/ ranking of research institutions, exploring the growth and change of disciplines and research topics over time, geographic research networks.
%\section{Discussion}
%\subsection{Lessons learned}
\section{Conclusion}
We presented the design and implementation of \textsc{Quoka}, an interactive atlas for exploring the research outputs of institutions around the world. The system supports sensemaking tasks related to understanding the creation and history of academic research, and can help to provide a more nuanced picture of the heterogeneity of research being conducted at different institutions. The \textsc{Quoka} service consists of a back-end data infrastructure and information retrieval index, combined with an interactive web-based interface.

Next steps will involve developing more sensemaking components for the \textsc{Quoka} system, including the integration of structured domain-based knowledge to support context-based search and to improve the usability and personalization of the system.

\section*{Acknowledgments}
The initial prototype of the \textsc{Quoka} atlas was developed during a generous visiting researcher opportunity provided to Benjamin Adams by the Curtin Institute of Computation from Oct.-Nov. 2019. This research was also supported by New Zealand Ministry of Business Innovation \& Employment, Grant/Award Number: UOAX1932. William Wallace helped with initial prototyping of the sandbox component.

\bibliographystyle{unsrtnat}
\bibliography{explorer} 

\end{document}